\documentclass[preprint,preprintnumbers,a4paper,11pt]{article}
\usepackage{pos}
\usepackage{graphicx}
\usepackage{caption}
\usepackage{hyperref}
\usepackage{url}

\usepackage{eso-pic}
\newcommand{\preprint}[2]{
   \AddToShipoutPictureBG*{%
      \AtPageUpperLeft{%
        \hspace{0.8\paperwidth}%
        \raisebox{#1\baselineskip}{%
          \makebox[0pt][l]{\textnormal{#2}}
   }}}%
}

\newcommand\Sec[1]{Section~\ref{Sec:#1}}

\newcommand\Fig[1]{Fig.~\ref{fig:#1}}
\newcommand\Eq[1]{Eq.~(\ref{eq:#1})}

\newcommand{\beq}{\begin{equation}}
\newcommand{\eeq}{\end{equation}}

\newcommand{\beqs}{\begin{eqnarray}}
\newcommand{\eeqs}{\end{eqnarray}}

\title{Strongly coupled gauge theories towards physics beyond the Standard Model}

\author*[a]{Jong-Wan Lee}

\affiliation[a]{Particle Theory and Cosmology Group, Center for Theoretical Physics of Universe, Institute for Basic Science, 
Daejeon 34126, Korea}

\emailAdd{j.w.lee@ibs.re.kr}

\abstract{
Strongly coupled gauge theories provide an ultra-violet realization of new physics models for physics beyond the Standard Model of particle physics arising from composite dynamics. Depending on the gauge group and matter content, they are expected to exhibit interesting features and rich phenomenology, similar or dissimilar to QCD, of which model builders and phenomenologists can take advantage. Due to the non-perturbative nature of these theories, first principles lattice calculations are essential to test the validity of composite models and provide theoretical inputs, that are otherwise unattainable, to the experimental searches for new physics. 
In this contribution, we will review recent efforts in the non-perturbative lattice studies of strongly coupled gauge theories other than QCD in the context of physics beyond the Standard Model by focusing on technical developments and new results. 
}

\FullConference{The 40th International Symposium on Lattice Field Theory (Lattice 2023)\\
July 31st - August 4th, 2023\\
Fermi National Accelerator Laboratory\\}

\begin{document}
\preprint{-12}{CTPU-PTC-24-03}

\maketitle

\section{Introduction}

\begin{figure}
\begin{center}
\includegraphics[width=0.87\textwidth]{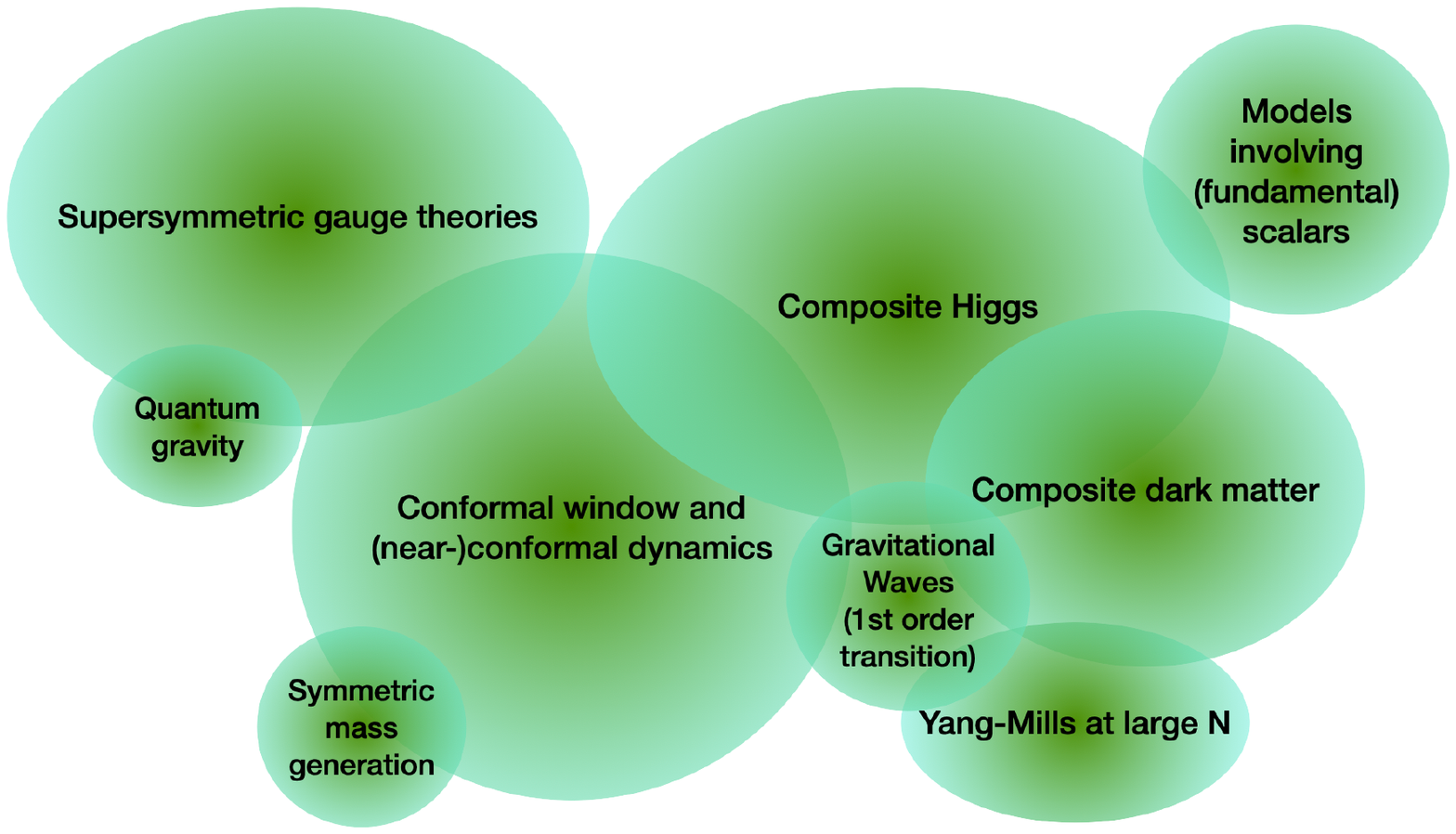}
\caption{%
\label{fig:latticebsm}%
Topics on lattice gauge theory for BSM physics presented at lattice conferences over the past three years. 
}
\end{center}
\end{figure}

The Standard Model of particle physics has been well established over the past few decades and has successfully passed a variety of experimental tests focusing on the electroweak sector. Yet, it should be considered as an effective description of physics below TeV scale, restricted to the visible sector of our universe. In particular, the existence of dark matter, matter-antimatter asymmetry, and non-zero neutrino mass cannot be explained by the Standard Model (SM). It is less obvious, but the SM also suffers from theoretical weaknesses, such as the naturalness/hierarchy problem, strong CP problem, and fermion mass hierarchy, which motivates us to search for a more fundamental description of physics in nature. The supersymmetric (SUSY) extensions of the SM have received much attention as they provide a solution to the hierarchy problem as well as an excellent candidate for the cold dark matter. Despite enormous experimental efforts to find SUSY particles, however, no evidence has yet been found, leading to numerous alternatives being proposed. Among them, BSM models based on new strongly coupled gauge theories (SCGTs) stand out for the following reasons. First of all, as witnessed by theoretical and experimental studies of QCD for the last half a century, a rich phenomenology can emerge at low energy as a result of confining and chiral symmetry breaking dynamics. In addition, some novel features of strongly interacting gauge theories, not yet explored, may have potential impacts on the BSM physics. Due to the strongly coupled nature of SCGTs, non-perturbative lattice calculations are crucial not only to test the validity of the models, but also to provide theoretical inputs necessary for the model building. 
Recent research activities along this line in the lattice community are summarized in \Fig{latticebsm}, where the number of contributions to each topic presented at lattice conferences over the past three years is indicated by the size of the bubble.

The last puzzle piece of the SM, the Higgs boson, was discovered by the ATLAS and CMS experiments at the large hadron collider (LHC) in 2012 \cite{ATLAS:2012yve,CMS:2012qbp}. As a direct evidence of new physics beyond the SM, experimental searches for new particle resonances at the LHC are still an active area of research, but not that successful at a few TeV scale or below. 
This provides a strong constraint on the BSM model buildings. 
In particular, if the Higgs is assumed to be a composite particle, such a large separation between the Higgs mass and the new physics scale can be addressed in two compelling scenarios, pseudo Nambu Goldstone Boson (pNGB) Higgs or dilaton Higgs. In the former, the spontaneous breaking of the chiral symmetry in the hyper-color gauge theory gives rise to light pNGBs, like pions in QCD, and part of them serve as the Higgs doublets in the SM. In modern pNGB Higgs models, the top quark is also considered to be (partially) composite, where its large mass can be obtained by the linear mixing with the baryonic operator composed of fermionic constituents transforming under one or multiple representations of the hyper color group. In the dilaton Higgs model, a light dilaton associated with the spontaneously broken approximate scale symmetry replaces the Higgs particle. Interestingly, nearly conformal dynamics is ubiquitous in composite Higgs scenarios. Because QCD with sufficiently large numbers of flavors are expected to be infra-red (IR) conformal, it has also been a longstanding theme in the lattice community of its own theoretical term. 

From various astronomical observations and the standard model of cosmology we learned that the mass budget of ordinary and dark matter in our universe is in the same order of magnitude, while the interactions between them are extremely weak. 
Assuming that dark matter has a particle origin, such observed features can be addressed by modeling dark matter as a composite particle of SCGTs in the dark sector. 
In particular, it is possible to make the composite particle arising from dark gauge dynamics neutral to the SM interactions even if the constituents in the underlying dark gauge theories are not. 
Furthermore, a rather strong self-interaction can easily be accommodated, and thus it is possible to naturally explain certain astronomical observations at small scales that are otherwise difficult to be addressed \cite{Tulin:2017ara}. Last but not least, the dark confinement transition could be first-order and may provide a potential source for the stochastic gravitational waves accessible to current and future gravitational-wave detectors. 

In this contribution we concentrate on the recent lattice studies of non-supersymmetric gauge theories coupled to vector-like fermions, towards composite Higgs, top-partial compositeness in \Sec{ch}, and composite dark matter in \Sec{cdm}, followed by the discussion on the conformal window and (near)-conformal dynamics in \Sec{cw}. 
Other topics, including supersymmetric gauge theories, gauge theories with elementary scalars, and symmetric mass generation, will also be briefly discussed in \Sec{others}. 

\section{Conformal window and (nearly) conformal dynamics}
\label{Sec:cw}

We first introduce the concept of conformal window in non-abelian gauge theories with $N_f$ flavors of massless fermion, and report recent results on its determination using perturbative and non-perturbative approaches. 
By focusing on $N_f=8$ $SU(3)$ gauge theory, we also present the numerical lattice results of low-lying spectrum including a light scalar, and discuss the EFT analysis of the results. 

\subsection{Conformal window}

Renormalization group (RG) equations describe the evolution of a physical system with respect to the energy scale, 
and have been used to understand how long-distance or infra-red (IR) dynamics emerge from short-distance or ultra-violet (UV) dynamics. 
In particular, the RG beta function $\beta(\alpha)=\frac{\partial \alpha}{\partial \ln \mu}$ of quantum field theory encodes the change of the coupling constant $\alpha$ with the renormalized scale $\mu$. 
If we restrict ourselves to non-abelian gauge theories coupled to $N_f$ massless fermions with a coupling constant $g$, 
the perturbative calculation at one-loop immediately finds $\beta(g)<0$ for $N_f < N_f^{\rm AF}$ and thus yields asymptotic freedom with a gaussian fixed point at $g=0$ \cite{Gross:1973id,Politzer:1973fx}. 
While theories with sufficiently small $N_f$, including Yang-Mills, are supposed to be confined and chirally broken, 
theories with $N_f$ just below $N_f^{\rm AF}$ (assuming fractional numbers of $N_f$) develop a perturbative IR fixed point at two-loop order \cite{Caswell:1974gg}, called {\it Banks-Zaks} (BZ) fixed point. 
Without loss of generality, asymptotic free gauge theories are therefore further divided into two distinct phases 
in which the IR features are completely different. A {\it conformal window} (CW) is defined as the range of $N_f$ over which 
the theories are IR conformal, i.e. $N_f^{\rm cr}< N_f< N_f^{\rm AF}$.

At the lower end of the conformal window theories are supposed to be strongly interacting, and thus it is  notoriously difficult to study their IR nature using the conventional perturbative approaches. 
Nonperturbative lattice calculations must be suitable for this purpose, but are still challenging, because, in contrast to QCD-like theories, the characteristic length scales of such theories are (infinitely) large. 
Furthermore, the nature of the zero-temperature quantum phase transition 
between the IR conformal and chirally broken phases 
is still largely unknown: it has been conjectured to be an infinite-order \cite{Kaplan:2009kr} or weak first-order \cite{Gorbenko:2018ncu}.  
Despite these difficulties, 
potentially attractive features that are not yet fully explored, such as the existence of a light scalar, large-scale separation, and the appearance of novel IR phases, keep motivating us to put more efforts to understand the (nearly) conformal dynamics using both perturbative and non-perturbative methods. 

The perturbative analysis of the conformal window can be used to make an estimate of $N_f^{\rm cr}$ by imposing that the coupling at the BZ fixed point runs to infinity, which can be improved by adding higher loop-order corrections to the beta function, $\beta(\alpha)=-2\alpha \sum_{\ell =1}^\infty b_\ell \left(\frac{\alpha}{4\pi}\right)^\ell$, e.g. see Ref.~\cite{Pica:2010xq}. 
However, this approach not only becomes scheme-dependent for $\ell \geq 3$, but also leads to serious convergence problems when the $5$th order results in the $\overline{\rm MS}$ scheme, the highest loop-order known to us so far, are concerned \cite{Ryttov:2016ner}. 
An alternative scheme-independent expansion in $\Delta_{N_f}=N_f^{\rm AF}-N_f^{\rm IR}$, called conformal or BZ expansion, has been considered in the context of (asymptotically free) IR conformal gauge theories \cite{Banks:1981nn,Ryttov:2016hdp} and, for the past few years, extensively used to compute certain physical observables at an IR fixed point by Ryttov and Shrock \cite{Ryttov:2016asb,Ryttov:2016hal,Ryttov:2017toz,Ryttov:2017kmx,Ryttov:2017dhd,Gracey:2018oym,Ryttov:2018uue,Ryttov:2020scx}. 
A particularly interesting quantity is the anomalous dimension of a fermion condensate
\begin{equation}
    \gamma_{\bar{\psi}\psi,\,{\rm IR}}(\Delta_{N_f})=\sum_{j=1}^{\infty} c_j(\Delta_{N_f})^j. 
    \label{eq:ga_conf}
\end{equation}
The coefficients $c_j$ are determined from the $(j+1)$- and $j$-loop order results of $\beta(\alpha)$ and $\gamma(\alpha)$: after expanding $\alpha_{\rm IR}$ with respect to $\Delta_{N_f}$, one uses the fact that $\beta(\alpha_{\rm IR})=0$ at each order in $\Delta_{N_f}$. 

While the disappearance of the IR fixed point could be direct evidence of conformality lost, the most relevant operator associated with the chiral phase transition might be the fermion condensate. 
In fact, various analytical studies, such as Schwinger-Dyson analysis of the gap equation \cite{Cohen:1988sq} and the chiral transition through the annihilation of IR and UV fixed points \cite{Kaplan:2009kr}, strongly suggest that the following critical condition is responsible for the loss of conformality:
\begin{equation}
    \gamma_{\bar{\psi}\psi}\equiv 1,~{\rm or~equivalently},~\gamma_{\bar{\psi}\psi}(2-\gamma_{\bar{\psi}\psi})\equiv 1.
    \label{eq:crit_con}
\end{equation}
Combining Eqs.~(\ref{eq:ga_conf}) and (\ref{eq:crit_con}), one can determine the lower end of the conformal window in a scheme-independent way \cite{Kim:2020yvr,Lee:2020ihn}. 
Note that at finite order in $\Delta_{N_f}$ the two critical conditions in \Eq{crit_con} generally give rise to different values of $N_f^{\rm cr}$, and the quadratic form has been shown a better convergence than the linear \cite{Lee:2020ihn}.\footnote{
Recently, a similar study for supersymmetric gauge theory has been reported in Ref.~\cite{Ryttov:2023xny}, where the authors find that the linear form of the critical condition shows a better convergence. 
}
After taking account of potential systematic uncertainties arising from the truncation in the BZ expansion as well as from the difference between two critical conditions at finite order, the sill of the CW in many-flavor QCD has been estimated to be $N_f^{\rm cr}=9.79^{+0.94}_{-0.82}{}^{+0.31}_{-0.36}$ \cite{Lee:2020ihn}. 
This approach has also been applied to non-abelian gauge theories coupled to fermions in multiple representations, in particular two-representation theories relevant to composite Higgs and top-partial compositeness \cite{Kim:2020yvr}. 
Very recently, another scheme-independent approach was proposed, 
where $N_f^{\rm cr}$ of $N_f$-flavor QCD has been estimated by computing $f_\pi/m_V$ and $f_V/m_V$ using (p)NRQCD in the conformal window and matching them to the non-perturbative lattice results for $N_f=[2,\,10]$, which finds a somewhat larger value of $N_f^{\rm cr}\sim 12$ or $13$ \cite{Chung:2023mgr}.

Nonperturbative lattice calculations have been playing a crucial role in the understanding of the IR properties of asymptotically free gauge theories. Among various approaches, recently, much efforts have been devoted to a direct investigation of the existence of an IR fixed point by calculating the non-perturbative RG beta functions, with the help of the gradient flow (GF) method yielding a renormalized coupling \cite{Luscher:2010iy}, either in discrete \cite{Fodor:2012td} or continuous definition \cite{Fodor:2017die,Hasenfratz:2019hpg}. 
A GF renormalized coupling is defined as
\beq
g_{\rm GF}^2 (t) = \mathcal{N} t^2\langle E(t)\rangle,
\eeq
where $E(t)$ is the energy density at gradient flow time $t$ and $\mathcal{N}$ is the normalization factor. 
At finite volume, $L^4$, the extent $L$ can be used to define the RG scale $\mu=1/L$ if $c=\sqrt{8t}/L$ is fixed. 
Similar to the step scaling function based on Schrodinger functional \cite{Luscher:1992an}, one can then obtain a discrete $\beta$-function by calculating the difference between $g_{\rm GF}^2(sL)$ and $g_{\rm GF}(L)$ for a scale change $s$, i.e. $\beta_s(g^2;L) = (g^2(sL)-g^2(L))/\log(s^2)$. The continuum limit can be achieved by repeating the calculations for a set of volumes and taking the infinite volume limit on the results. The results for $N_f$-flavor QCD can be found in Refs.~\cite{Hasenfratz:2019dpr,Hasenfratz:2020ess,Hasenfratz:2022yws,Hasenfratz:2022zsa}, e.g. see Fig.~1 in \cite{Hasenfratz:2022zsa}, where the $N_f=12$ theory develops an IR fixed point at $g_{\rm GF}\sim 6$ in the $c=0.3$ scheme. 

The continuous RG $\beta$-function, on the other hand, is defined by taking the derivative of $g_{\rm GF}^2$ with the flow time $t$ itself, $\beta(g_{\rm GF}^2) = -t d g_{\rm GF}^2 / dt$ \cite{Fodor:2017die,Hasenfratz:2019hpg}. 
Such $\beta$-functions obtained from different bare actions result in different RG trajectories, but converge to a continuum one in the large flow time limit as irrelevant terms in the lattice definition of $E(t)$ fade away.\footnote{
At finite fermion mass $E(t)$ would involve a relevant operator and the corresponding RG trajectory flows away from the critical surface. 
In such cases, the analysis is complicated by additional chiral extrapolations, e.g. $am\rightarrow 0$.\cite{Fodor:2017die}
} 
In practice, one first takes an infinite volume limit at fixed $t/a^2$ as such the continuous RG is only valid in infinite volume. Next, an infinite flow time extrapolation is carried out at every $g_{\rm GF}^2$. 
Assuming that the theory exhibits an IR-conformal fixed point, the continuous RG transformation can also be used to compute an anomalous dimension of a local gauge-invariant operator \cite{Carosso:2018bmz}. 
More specifically, one considers the $2$-point correlation function of a flowed operator $\mathcal{O}_t(x_0)$ with an unflowed one $\mathcal{O}(0)$, $\langle \mathcal{O}(0)\mathcal{O}_t(x_0)\rangle$. 
After accounting for the wavefunction renormalization by taking a ratio with the conserved current, e.g. vector current, the anomalous dimension of the desired operator $\mathcal{O}$ can be extracted from the logarithmic derivative at $x_0 \gg \sqrt{8t}$. 

\begin{figure}
\begin{center}
\includegraphics[width=0.47\textwidth]{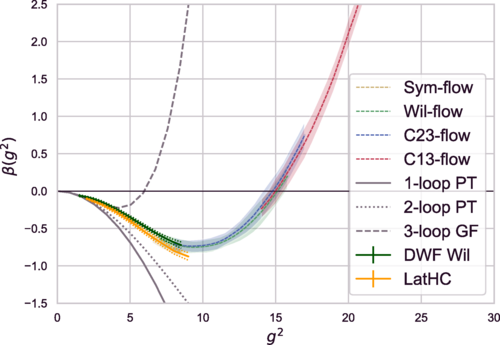}
\hspace{3.0mm}
\includegraphics[width=0.445\textwidth]{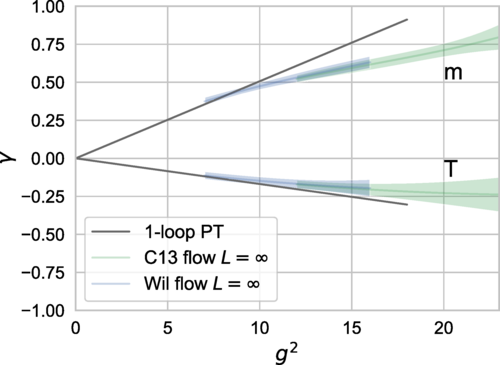}
\caption{%
\label{fig:su3_10F}%
(left) The continuous renormalization group beta function and (right) the anomalous dimension of mass operator (and that of tensor density) in the $SU(3)$ gauge theory with $N_f=10$ flavors from Ref.~\cite{Hasenfratz:2023wbr}. 
}
\end{center}
\end{figure}

The continuous RG method has been applied to the $SU(3)$ gauge theories with $N_f=2$, $12$ fundamental Dirac flavors \cite{Hasenfratz:2019puu} in which the $N_f=12$ theory exhibits an IR fixed point, being consistent with the results from the GF step-scaling function approach \cite{Hasenfratz:2019dpr}. 
Very recently, it has also been used to compute the continuous RG beta function and the mass anomalous dimension of the $SU(3)$ theory with $N_f=10$ flavors \cite{Hasenfratz:2023wbr} --- the most controversial theory in the context of conformal window. There, the authors have introduced heavy Pauli-Villars (PV) bosons to remove UV fluctuations \cite{Hasenfratz:2021zsl} and have been able to reach the GF coupling to $g_{\rm GF}\sim 20$ for the first time. 
As shown in \Fig{su3_10F}, the results strongly suggest that $N_f=10$ theory is IR conformal, in good agreement with the perturbative prediction discussed above.  
At the IR fixed point $g_{\rm GF}\simeq 15$, the resulting value of the mass anomalous dimension, $\gamma_m \equiv \gamma_{\bar{\psi}\psi,\,{\rm IR}} \simeq 0.6$, is also consistent with the perturbative value of $\gamma_{{\bar{\psi}\psi},\,\rm IR} (\Delta_{N_f}^4) = 0.615$ obtained from the scheme-independent conformal expansion 
\cite{Ryttov:2016asb}, but is slightly larger than the one $\gamma_m=0.47(5)$ extracted from the $4+6$ mass-split model~\cite{LatticeStrongDynamics:2020uwo}. 
A similar study for the $N_f=8$ $SU(3)$ theory is under way, where the conclusion from preliminary results presented by C.~Peterson et al. in this conference is inconclusive, but seems to suggest that the theory is just below the conformal window: no sign of an IR fixed point to $g_{\rm GF}^2\simeq 22$, while the BKT scaling being preferred. 

\subsection{Near-conformal dynamics}

Nearly conformal gauge theories are relevant to many extensions of the SM, especially composite Higgs and top partial compositeness. 
The most striking feature of such theories, revealed recently from various lattice calculations, is the emergence of a light $0^{++}$ scalar in the low-lying spectrum which could be identified as the dilaton. This has triggered a revival of dilaton effective field theory (e.g. see Ref.~\cite{coleman}) and has sparked much interest in certain composite Higgs scenarios in which the light scalar plays the role of Higgs \cite{Goldberger:2007zk}. 
Two most well-known prototypes, extensively studied on the lattice, are the $SU(3)$ gauge theories coupled to $N_f=8$ fundamental and $N_f=2$ symmetric (sextet) Dirac fermions. The recent numerical results for the former are found in Refs.~\cite{LatticeStrongDynamics:2018hun,LatticeStrongDynamicsLSD:2021gmp,LatticeStrongDynamics:2023bqp} by Lattice Strong Dynamics (LSD) collaboration and in Ref.~\cite{LatKMI:2016xxi} by LatKMI, and those for the latter are in Ref.~\cite{Fodor:2016pls} by Fodor et. al. (see also Ref.~\cite{Hansen:2017ejh}). 
Although it is not fully confirmed yet, these models are widely believed to be slightly outside of the conformal window. 

Several approaches have been proposed to describe the low-energy spectrum of near-conformal gauge theories containing a light scalar, as light as pNGBs, e.g. see Refs.~\cite{Witzel:2019jbe,Drach:2020qpj} and references therein. Among them, the effective field theory approaches, which interpret the light iso-singlet scalar as a dilaton field associated with the spontaneous  breaking of scale symmetry, have been developed over many years, and applied to analyze lattice data in a series of papers by Golterman, Shamir \cite{Golterman:2016lsd,Golterman:2018mfm,Golterman:2020tdq,Golterman:2020utm} (and Freeman \cite{Freeman:2023ket}) and Appelquist, Ingoldby, Piai \cite{Appelquist:2017wcg,Appelquist:2017vyy,Appelquist:2019lgk,Appelquist:2022mjb}, independently. 
In the former approach, the authors considered the Veneziano limit such that the number of flavors can be treated as a continuous parameter, $x_f=N_f/N_c$, and utilized it to construct their dilaton chiral perturbation theory (dChPT) by performing a systematic expansion in terms of $|x_f-x_f^{\rm cr}| \ll 1$ which controls the breaking of scale or dilatation symmetry, in addition to the usual chiral expansion. 
One of the main lessons of their work is that in the large-mass region hadron masses and decay constants behave according to the hyperscaling relation, $m_f^{1/(1+\gamma_m)}$, as in the mass-deformed conformal theory \cite{DelDebbio:2010ze}, while in the small mass region the typical chiral behavior is restored as the dilaton decouples from the low-enery dynamics. They have applied dChPT to the lattice data of the $N_f=8$ theory, and concluded that the LSD data in Ref.~\cite{LatticeStrongDynamics:2018hun} are likely to be in the large-mass region where LO dChPT is applicable \cite{Golterman:2018mfm,Golterman:2020tdq}, but the LatKMI data in Ref.~\cite{LatKMI:2016xxi}, obtained with larger fermion masses, seem to lie outside the dChPT range. \cite{Golterman:2020utm,Freeman:2023ket}. 
See also Ref.~\cite{Fodor:2019vmw} for the case of the $N_f=2$ sextet model.

In the latter approach led by Appelquist, Ingoldby and Piai, the dilaton potential which explicitly breaks the conformal symmetry has been assumed to be a rather generic form, having a minimum at $f_d >0$ and satisfying $m_d^2\ll (4\pi f_d)^2$, but leaving the details to be determined by the lattice data. After introducing a real scalar field $\chi$ for the dilaton and incorporating the scaling behavior to the chiral Lagrangian, they defined the dilaton EFT (dEFT) Lagrangian at leading order \cite{Appelquist:2019lgk}
\beq
\mathcal{L}_{\rm LO} = \frac{1}{2}\partial_\mu \chi \partial^\mu \chi + 
\frac{f_\pi^2}{4}\left(\frac{\chi}{f_d}\right)^2 {\rm Tr} \left[
\partial_\mu \Sigma (\partial^\mu \Sigma)^\dagger
\right]
+\frac{m_\pi^2 f_\pi^2}{4}\left(\frac{\chi}{f_d}\right)^y
{\rm Tr} \left[\Sigma+\Sigma^\dagger \right] -V_\Delta (\chi),
\label{eq:deft_lo}
\eeq
where the tree-level dilaton potential $V_\Delta(\chi)$ has the form
\beq
V_\Delta(\chi) \equiv \frac{m_d^2 \chi^4}{4(4-\Delta)f_d^2}
\left[
1-\frac{4}{\Delta}\left(\frac{\chi}{f_d}\right)^{\Delta-4}
\right].
\label{eq:d_potential}
\eeq
By varying the scaling dimension $\Delta$, this potential interpolates several known forms, from the usual Higgs potential with $\Delta=2$ to the one arising from the marginal deformation of the conformal theory in the limit $\Delta\rightarrow 4$ \cite{Goldberger:2007zk}. 
The effective Lagrangian in \Eq{deft_lo} with the choice of $\Delta\rightarrow 4$ is essentially identical to the LO dChPT in Ref.~\cite{Golterman:2016lsd}.  
For further discussion of the interpretation of $\Delta$, see Ref.~\cite{Ingoldby:2023mtf}

\begin{figure}
\begin{center}
\includegraphics[width=0.81\textwidth]{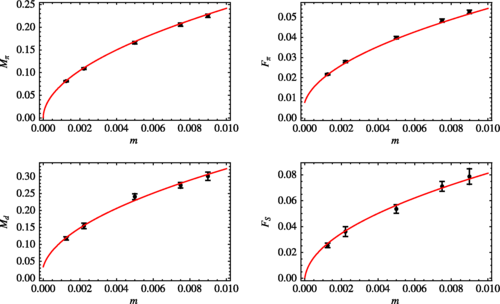}
\includegraphics[width=0.47\textwidth]{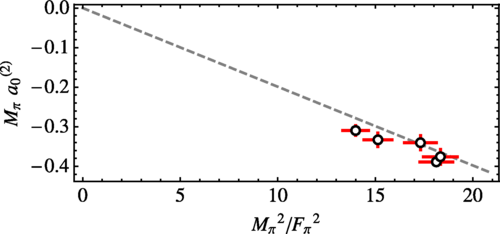}
\caption{%
\label{fig:lsd_dilaton}%
Dilaton EFT fits to the lattice results of the $SU(3)$ gauge theory with $N_f=8$ fundamental flavors. The plots are taken from  Ref.~\cite{LSD:2023uzj}. 
}
\end{center}
\end{figure}

The LSD collaboration recently measured new observables in the $N_f=8$ theory: the s-wave scattering length of Goldstone bosons in the maximal-isospin channel \cite{LatticeStrongDynamicsLSD:2021gmp} and the scalar decay constants \cite{LatticeStrongDynamics:2023bqp}. In Ref.~\cite{LatticeStrongDynamics:2023bqp}, the authors also updated their previous results in Ref.~\cite{LatticeStrongDynamics:2018hun} by reducing the systematic errors substantially with several technical improvements, including the infinite volume extrapolation and the vacuum subtraction for the flavor-singlet scalar correlator. Combining all these lattice data, they performed a global fit using the dilaton EFT \cite{LSD:2023uzj}. The fit results are shown in \Fig{lsd_dilaton}, where LO dEFT describes the lattice data very well. To investigate the possibility that the theory lies within the conformal window, they also performed another fit to the same data using a simple hyperscaling relation and found that the quality was slightly lower.

\section{Composite Higgs and top partial compositeness}
\label{Sec:ch}

In this section, we review the recent progress of numerical studies on the lattice models relevant to the UV realization of composite pNGB Higgs arising from new strong interaction at a few TeV scale. 
Of our particular interests are in the $SU(4)$ and $Sp(4)$ hypercolor gauge theories coupled to fermions in two different representations, where one can naturally accommodate the top partial compositeness via the linear mixing with a chimera baryon in the new strong sector---top partner---carrying the same SM quantum numbers of top quark \cite{Ferretti:2013kya,Ferretti:2016upr,Belyaev:2016ftv}. 
We also discuss the composite Higgs model based on $SU(2)$ gauge theory with two fundamental flavors, and related lattice calculations. Although not covered in this review, we should note that the aforementioned eight-flavor $SU(3)$ gauge theory can also be used to construct another composite pNGB Higgs scenarios with in the framework of dilaton effective field theory \cite{Appelquist:2020bqj,Appelquist:2022qgl}. 

\subsection{$SU(4)$ gauge theory}

The $SU(4)$ gauge theory that serves as a UV completion of composite Higgs models proposed in Ref.~\cite{Ferretti:2013kya} contains either $3$ fundamental (F) Dirac and $5$ two-index antisymmetric (AS) Majorana fermions or $4$ F Dirac and $6$ AS Majorana fermions, which is expected to be slightly below the conformal window. On the lattice the theory with $2$ F and $2$ AS (sextet) Dirac fermions, the ``$2+2$ $SU(4)$ model'', has instead been extensively studied by TACoS collaboration for the past years: from the spectra of mesons composed of hyperquarks in one of the representations \cite{Ayyar:2017qdf}, to the spectra of baryons composed of hyperquarks in the fundamental, sextet, and both representations \cite{Ayyar:2018zuk}, to the chimera baryon matrix elements entering to the top-Yukawa coupling \cite{Ayyar:2018glg}, to the low-energy constants entering to the Higgs potential \cite{Ayyar:2019exp}, and to the contribution of the hypercolor theory to the $S$ parameter in terms of the vacuum alignment parameter \cite{Golterman:2020pyx}. 
This prototype model turned out to be QCD-like, as expected from the fact that it is deep inside the chirally broken phase, e.g. see the black dot in the left panel of \Fig{su4_4f4as_cw}, and is likely ruled out as a realistic model for composite-Higgs and top partial compositeness. 
Nevertheless, a series of these calculations paved a way to non-perturbative lattice studies of a mixed-representation gauge theory. 

\begin{figure}
\begin{center}
\includegraphics[width=0.47\textwidth]{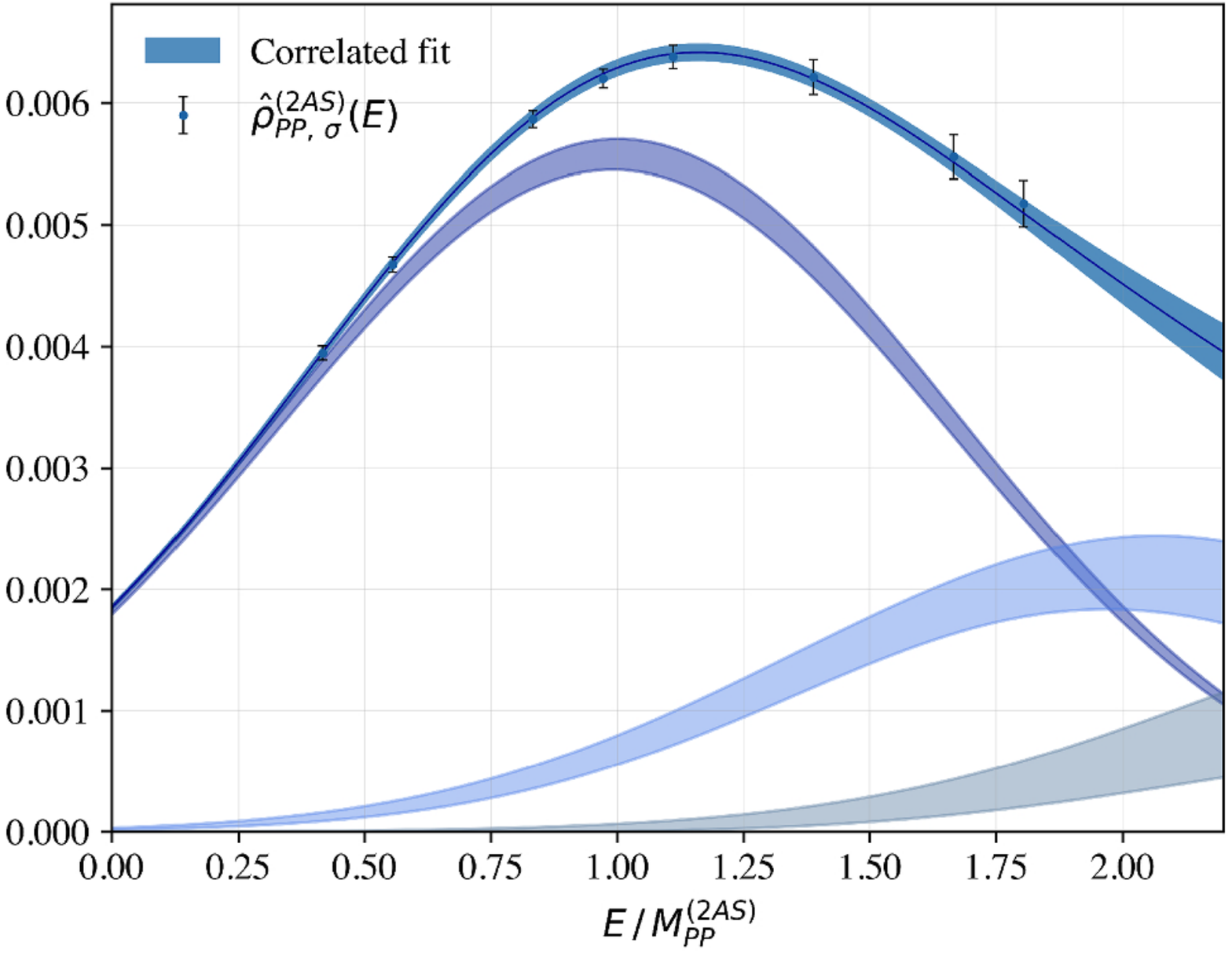}
\includegraphics[width=0.47\textwidth]{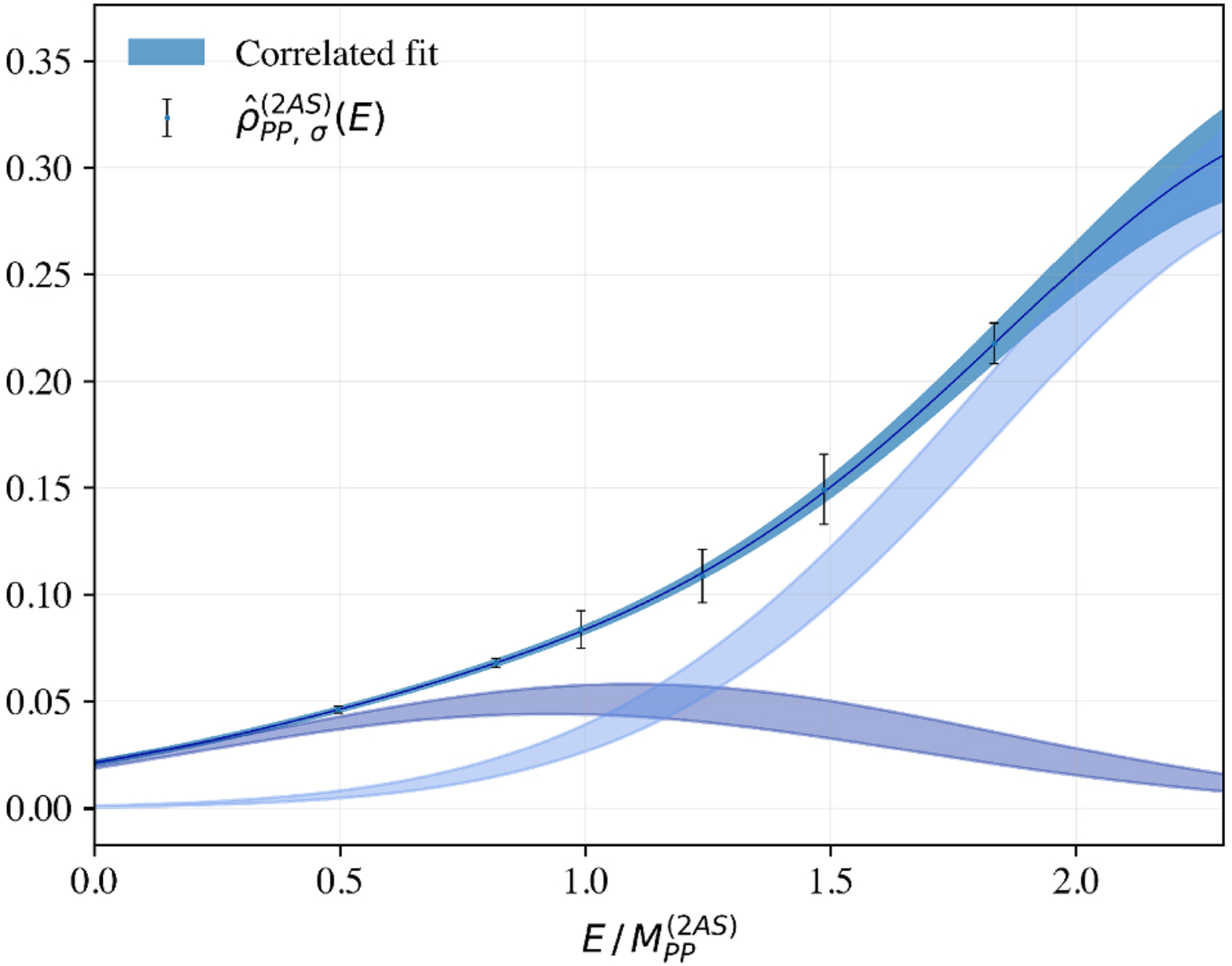}
\caption{%
\label{fig:spectral_density_su4}%
Examples of gaussian-smeared spectral densities, and Gaussian fits, extracted from a pseudoscalar-pseudoscalar correlation comprised of antisymmetric fermions in the $2+2$ $SU(4)$ gauge theory~\cite{DelDebbio:2022qgu}. The results in the left and right panels have been obtained with smeared and local interpolating fields, respectively. 
}
\end{center}
\end{figure}

The same model has also been studied on the lattice with a focus on the technical developments which might play a crucial role in a non-perturbative understanding of the spectrum of the multi-representation theories \cite{Cossu:2019hse,DelDebbio:2022qgu}. 
In particular, the possible excited states of a pseudoscalar meson $\Pi$ in the antisymmetric sector start with $\Pi \Pi$, due to the triviality of $G$-parity, while in the fundamental sector they start with $\pi \pi \pi$ in analogy with QCD. The extraction of the ground state of the AS pseudoscalar meson could be further complicated by a potentially shallow mass gap with the excited states composed of constituents in both representations, such as $\Pi \pi \pi$, $\Pi \pi \pi \pi$, and so on, if the F pseudoscalar is much lighter than the AS one. The authors employed and tested the recently developed spectral density method in Ref.~\cite{Hansen:2019idp} to tackle these challenges, where they found that the results are compatible to other established methods and operator smearing is essential in this endeavor. \Fig{spectral_density_su4} shows two examples of correlated fits of the gaussian-smeared spectral densities, where the excited state suppression is clearly seen when the smeared operators have been used to compute the correlation function. 

The other lattice $SU(4)$ model considered in the recent publication of Ref.~\cite{Hasenfratz:2023sqa} contains $4$ F and $4$ AS Dirac fermions, the ``4+4 $SU(4)$ model''. In contrast to the $2+2$ $SU(4)$ model denoted by black square in the left panel of \Fig{su4_4f4as_cw}, this model, denoted by open circle, is expected to be in the sill of the conformal window, as suggested by the analytical estimations including the ones discussed the previous section. 
The phenomenological composite Higgs models, denoted by blue circle and red diamond in the same figure, could be reached by making a subset of fermions to be infinitely heavy. As shown in the right panel of \Fig{su4_4f4as_cw}, it has been found that the continuous RG beta function reveals a stable IR fixed point at $g_{\rm GF}^2\simeq 15$, strongly indicating that the $4+4$ $SU(4)$ model is IR conformal~\cite{Hasenfratz:2023sqa}. At the IR fixed point, the mass anomalous dimensions have been calculated to be $\gamma_m^{(4)}\simeq 0.75$ and $\gamma_m^{(6)}\simeq 1.0$, which are in good agreement with the perturbative results reported in Ref.~\cite{Ryttov:2023uzc}. The authors also calculated the anomalous dimensions of top-partner chimera baryon operators, $\gamma_{\rm ch} \simeq 0.5$ being the largest, but it turned out to be much smaller than the phenomenologically preferred value of $\gamma_{\rm ch}=2$ as well as the one-loop predictions \cite{DeGrand:2015yna,BuarqueFranzosi:2019eee}. 

\begin{figure}
\begin{center}
\includegraphics[width=0.45\textwidth]{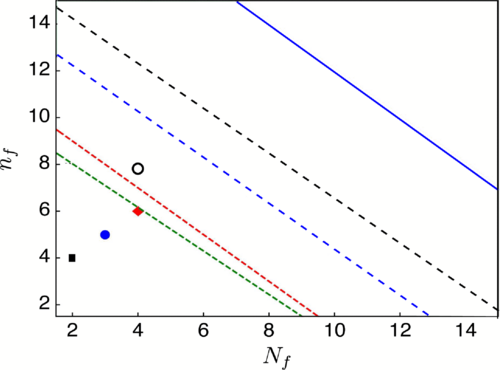}
\hspace{4.0mm}
\includegraphics[width=0.48\textwidth]{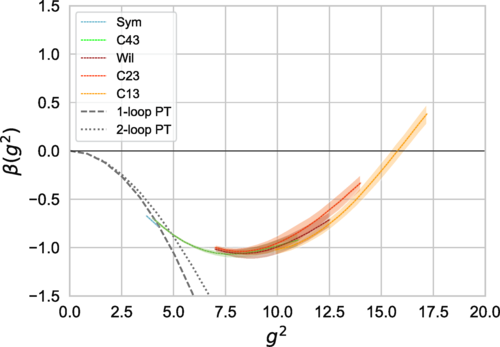}
\caption{%
\label{fig:su4_4f4as_cw}%
(left) The theory space of the $SU(4)$ gauge theories coupled to $N_f$ fundamental Dirac and $n_f$ antisymmetric Majorana fermions. The solid line corresponds to the upper bound of the conformal window above which asymptotic freedom is lost, while the dashed lines denote various analytical estimations for the lower bound.  (right) The continuous renormalization group beta function of the $4+4$ $SU(4)$ theory. The plots are taken from Ref.~\cite{Hasenfratz:2023sqa}.
}
\end{center}
\end{figure}

\subsection{$Sp(4)$ gauge theory}

A promising UV completion of composite Higgs and top partial compositeness proposed in Refs.~\cite{Ferretti:2013kya,Barnard:2013zea} is based on the $Sp(2N)$ gauge group and the coset of $\frac{SU(4)\times SU(6)}{Sp(4)\times SO(6)}$ arising from the (spontaneous) breaking of global flavor symmetry. One nice feature of this model is that it can be studied on the lattice with the exactly same global symmetry by introducing $N_f=2$ and $n_f=3$ Dirac fermions transforming in the fundamental and antisymmetric representations of the $Sp(4)$, respectively, without much technical difficulties \cite{Bennett:2022yfa}. 
The establishment of the $Sp(2N)$ lattice gauge theories with and without fermionic matter, including the model mentioned above, began in Ref.~\cite{Bennett:2017kga}, and the recent progress was summarized in a semi-review article of Ref.~\cite{Bennett:2023wjw}. 

\begin{figure}
\begin{center}
\includegraphics[width=0.89\textwidth]{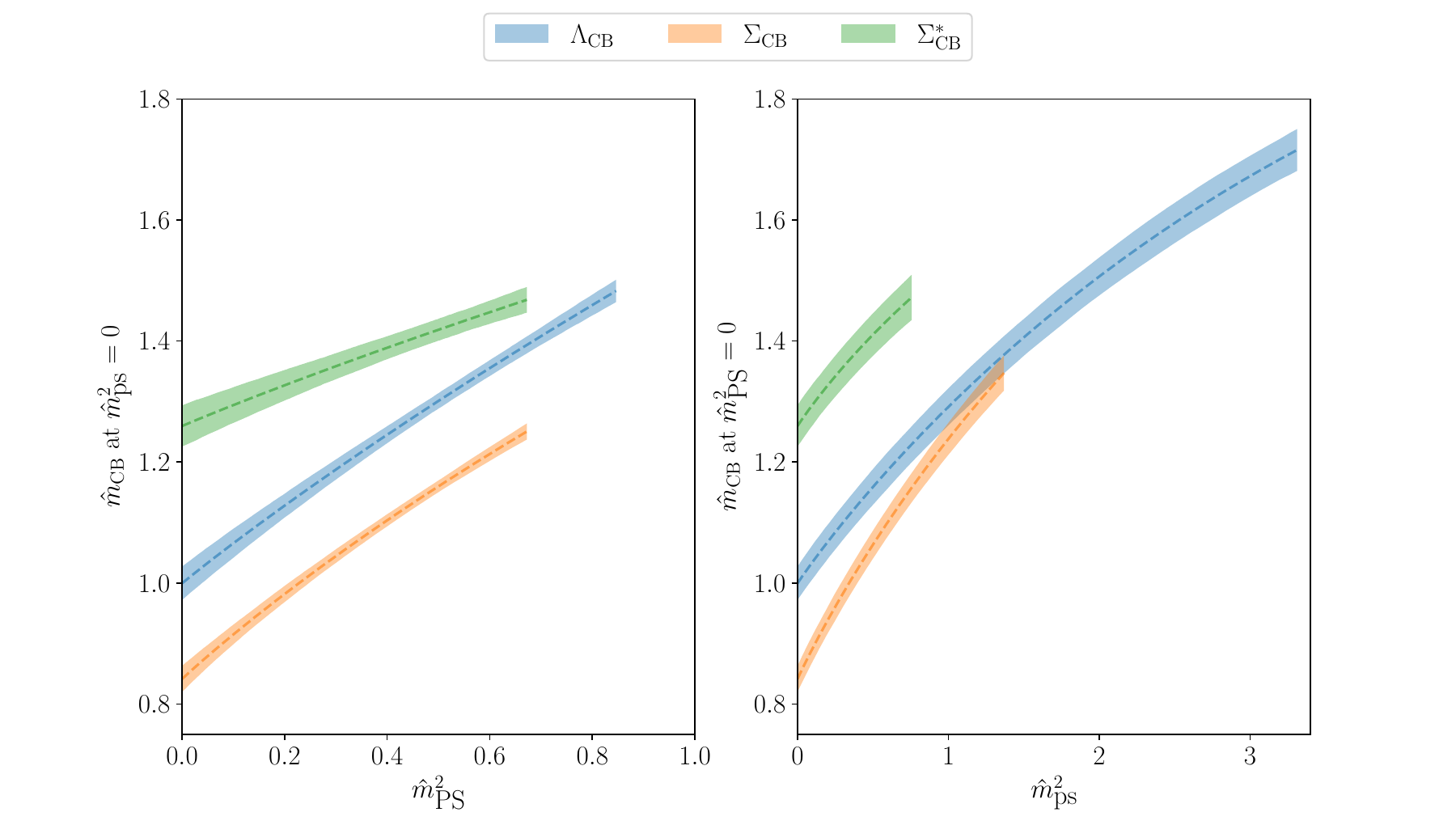}
\caption{%
\label{fig:cb_sp4}%
Masses of the lowest-mass chimera baryons in the continuum limit of quenched $Sp(4)$ gauge theory as a function of the pseudoscalar mass squared \cite{Bennett:2023mhh}. The mass of a pseudoscalar meson composed of either AS (left) or F fermions (right) has been extrapolated to zero beforehand. 
}
\end{center}
\end{figure}

\begin{figure}
\begin{center}
\includegraphics[width=0.775\textwidth]{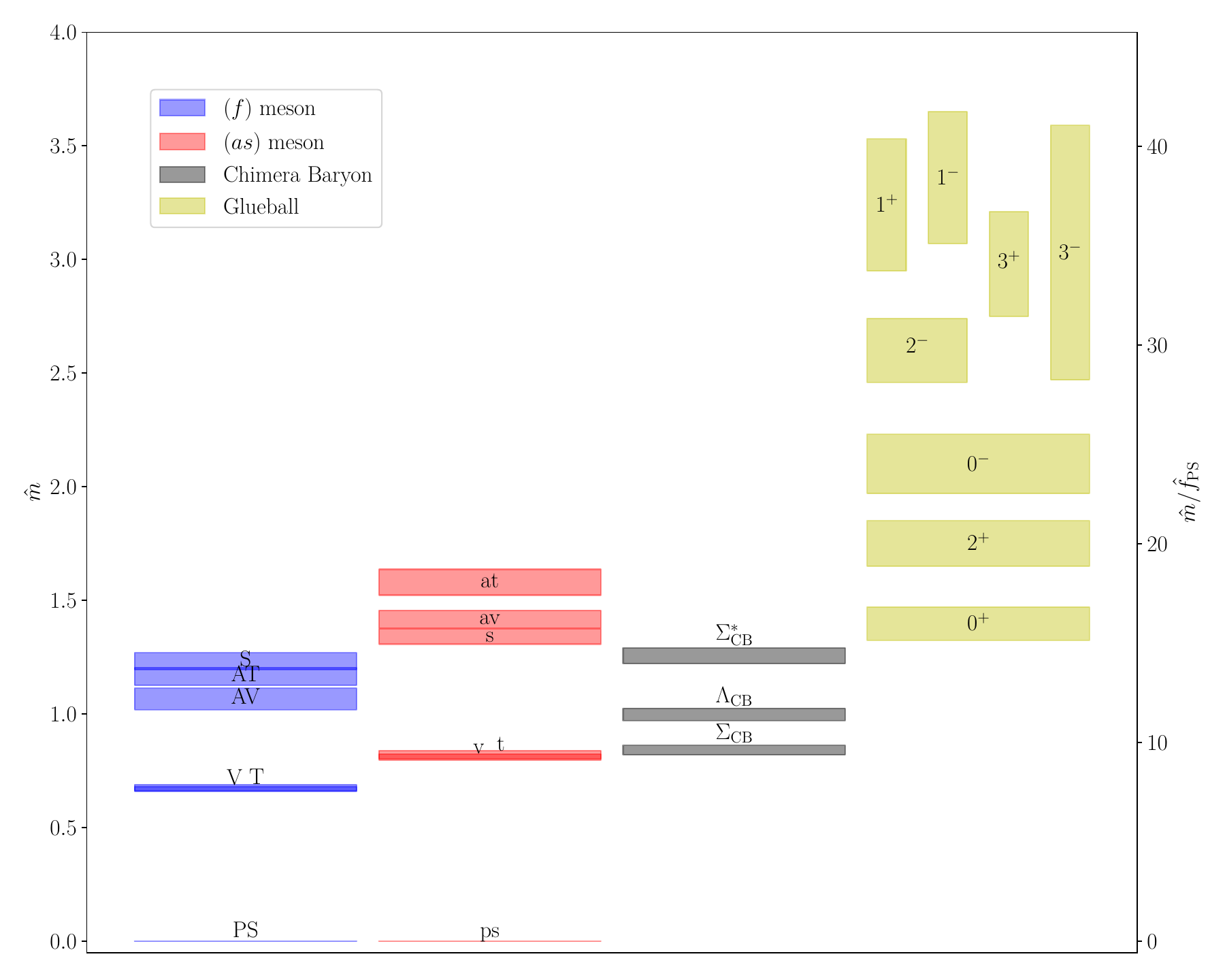}
\caption{%
\label{fig:quenched_spectra_sp4}%
Mass spectra of the quenched $Sp(4)$ gauge theory in the continuum and massless-hyperquark limits. The plot is taken from Ref.~\cite{Bennett:2023mhh}.
}
\end{center}
\end{figure}

Very recently the authors in Ref.~\cite{Bennett:2023mhh} carried out the measurements of the chimera baryon masses in the quenched limit of the $Sp(4)$ theory, where the continuum and massless extrapolated results in units of the gradient flow scale $w_0$ are shown in \Fig{cb_sp4}, e.g. $\hat{m} = m w_0$. Here, $\Lambda_{\rm CB}$, $\Sigma_{\rm CB}$, and $\Sigma^*_{\rm CB}$ denote the lowest-mass chimera baryons interpolated by the hypercolor-singlet operators, comprised of two F and one AS valence fermions, with the spin, parity and the irreducible representation of the fundamental sector, $(J^P,~F_{\rm irrep})= (\frac{1}{2}^+,~5),~(\frac{1}{2}^+,~10)$, and $(\frac{3}{2}^+,~10)$, respectively. In the small-mass region, one finds a clear mass hierarchy of $m_{\Sigma_{\rm CB}} < m_{\Lambda_{\rm CB}} < m_{\Sigma^*_{\rm CB}}$. Combining the previous results on the masses of mesons \cite{Bennett:2019cxd}, for which the valence fermion is either in the F or AS representation, and of glueballs in the various channels \cite{Bennett:2020qtj}, in \Fig{quenched_spectra_sp4} we present the complete quenched spectrum of the $Sp(4)$ theory in the massless limit. With a caveat of the quenched approximation, the lightest chimera baryon has the mass comparable to the vector meson composed of AS fermions. 
The other recent developments include the implementation of the symplectic gauge theories within the GRID framework \cite{Boyle:2015tjk} to set the stage needed to perform further extended studies towards the (near-)conformal dynamics \cite{Bennett:2023gbe}, and the measurements of the meson masses and decay constants for fermion matter content either in the fundamental, or the rank-$2$, antisymmetric or symmetric representations, treated in the quenched approximation of $Sp(2N)$ theories with $N=1,\,2,\,3,\,4$ and in the large-$N$ limit \cite{Bennett:2023qwx}. 

\subsection{$SU(2)$ gauge theory}

$SU(2)$ gauge theory coupled to $N_f=2$ fundamental fermions exhibits spontaneous symmetry breaking of the enhanced $SU(4)$ flavor symmetry and provides the minimalist UV completion of pNGB composite Higgs based on the $SU(4)/Sp(4)$ coset. Depending on how the electroweak symmetry is embedded, this theory can also realize the scenario in which the Higgs emerges as a massive excitation of the condensate \cite{Cacciapaglia:2014uja}. The most recent lattice studies have focused on the scattering of pseudoscalar mesons and the resonance properties either in the vector or singlet channel by employing improved Wilson actions both for the gauge and fermion sectors. In Ref.~\cite{Drach:2020wux}, the authors calculated the eigen-energies of the system with the same quantum number of vector resonances, $J^P=1^{-1}$, extracted the infinite-volume phase shift using L\"{u}scher's finite volume formalism. Using the Breit Wigner parametrization of the scattering amplitude, they determined the coupling, $g_{\rm VPP} = 7.8\pm 0.6$, not far from the QCD value of $g_{\rho\pi\pi}\simeq 6$, but slightly smaller than the one estimated from the KSRF relation, $g_{\rm VPP}^{\rm KSRF}=9.4\pm 0.6$.  

The scattering amplitude for the singlet channel has also been investigated in Ref.~\cite{Drach:2021uhl}. From two dynamical ensembles with different pion masses, the authors computed the lowest two energy levels at zero net momentum by solving the generalized eigenvalue problem with singlet and two-pion operators, and observed that the lowest states are well below the two-particle threshold. 
From the phase shift analysis using the L\"{u}scher's method, they concluded that the flavor-singlet scalar meson is a two-pion bound state in the explored region of fermion masses. They also found inconsistency between their results and the LO ChPT prediction, which indicates that the ensembles are in the mass range beyond which the LO ChPT is applicable. 

\section{Composite dark matter}
\label{Sec:cdm}

Dark matter makes up about $85\%$ of the total mass of our universe according to the standard $\Lambda$-CDM model of cosmology. 
Dark matter has only been evidenced by gravitational effects, yet it is widely believed to have its particle origin, and this has provided another theoretical motivation to extend the Standard Model of particle physics and sparked massive experimental efforts to search for signatures of particle dark matter. Given that the standard weakly interacting dark matter (WIMP) paradigm has been highly constrained by direct and indirect detection experiments, the strongly-coupled composite dark matter 
arising from the newly confining dynamics in the dark sector provides a compelling paradigm. Some common features of such scenarios are as follows:
\begin{itemize}
\item composite states such as mesons, baryons, and glueballs in QCD-like theories can be dark matter candidates,
\item stability is guaranteed by accidental symmetry, e.g. $U(1)_B$ baryon number symmetry, $G$-parity, etc,
\item interactions with SM particles are highly suppressed by symmetries in the current universe, but not in the early universe,
\item dark matter self-interactions are naturally accommodated, and could be strong enough to provide solutions to the small-scale structure problems,
\item the dynamical scale, $\Lambda_{\rm D}$, associated with the confinement can be used to develop an EFT description of the composite dark matter, and if the confinement transition is first order, it may produce a natural source of stochastic gravitational waves from the early universe.
\end{itemize}

Again, $SU(4)$ and $Sp(4)$ gauge theories have widely been used for the minimal UV models of composite dark matter, where the dark matter candidates are baryons in the former, known as the stealth dark matter, but they are pNGBs in the latter which serve as the strongly interacting massive particles (SIMPs). 
While most of the lattice results discussed in the context of composite Higgs can be recycled, the preferred parameter space and the phenomenologically useful observables are quite different, as we will discuss below. 

In the SIMP scenario, the dark matter relic abundance is determined thermally by the freeze-out of a number-changing $3\rightarrow 2$ self-annihilation process between dark matter particles, instead of $2\rightarrow 2$ annihilations into SM particles, and dark matter is in thermal equilibrium with the SM sector at the time of freeze-out \cite{Hochberg:2014dra}. This scenario typically predicts light dark matter with the mass in the sub-GeV. The minimal UV models realizing the SIMP paradigm are based on $Sp(2N)$ gauge theories coupled to $N_f=2$ Dirac fermions. The number changing process arises mainly from the $5$-point pNGBs interaction via the Wess-Zumino-Witten (WZW) anomaly term \cite{Hochberg:2014kqa}, which can further be generalized by including vector resonances or mass-splitting effects, e.g. see Ref.~\cite{Kulkarni:2022bvh} and references therein. Flavor-singlet pseudoscalar also plays an important role in dark matter phenomenology. For instance, it can be stable unless charged under the dark $U(1)'$ gauge group, and may participate in dark pion scattering. In this regard, the meson spectrum of two-flavor $Sp(4)$ theory has been studied on the lattice for the flavor-non-singlets \cite{Bennett:2019jzz}, and for the flavor-singlets with and without the iso-spin breaking \cite{Bennett:2023rsl}. Over the mass range of $0.65\lesssim m_{\rm PS}/m_{\rm V}\lesssim 0.9$, the vector and flavor-singlet mesons are not much heavier than pNGBs, similar to the two-flavor $SU(2)$ and $SU(3)$ gauge theories, which implies that these states would be long-lived, but should not be as long to contribute the current dark matter density, and can be observed in collider experiments. Another ongoing lattice calculation for the same theory concerns the scattering properties of dark pions using L\"uscher's method, where the preliminary results for the isospin-$2$ channel shows that the s-wave scattering length is negative across all ensembles considered \cite{Dengler:2023szi}. Using the constraints on the dark matter cross-section from astrophysical observations, the authors find that the mass of SIMP dark matter is predicted to be $m_{\rm DM} > 115\,{\rm MeV}$. 

The stealth dark matter is a model for composite dark matter built upon the $SU(N_D)$ gauge theory, with even $N_D\geq 4$, coupled to fundamental matter fields \cite{LSD:2014obp}. The lightest baryon provides a viable candidate for dark matter, where its stability is guaranteed by the $U(1)_B$ dark baryon number symmetry. Dark fermions are charged under the electroweak group of the SM in the vector-like representations, which allows a coupling to SM particles through the Higgs boson. Yet, the interaction of dark baryons with SM particles are largely prohibited: no dimension-$4$ interactions except the ones involving the Higgs boson, no dimension-$5$ magnetic dipole interaction thanks to the scalar nature of dark baryon, and no dimension-$6$ charge radius operators given that the custodial $SU(2)$ is preserved. The leading interaction with photons occurs through the electromagnetic polarizability of dimension-$7$, which is highly suppressed by the dark confinement scale $\Lambda_{\rm D}$. In the minimal $SU(4)$ gauge theory (and the $SU(3)$ for comparison), lattice calculations of the spectrum of composite particles, the effective Higgs coupling, and the dark baryon electromagnetic polarizability have been carried out on quenched configurations \cite{LSD:2014obp,Appelquist:2015zfa,Appelquist:2015yfa}. The numerical results combined with experimental constraints yield the dark baryon mass to be a few hundred GeV. 
The non-perturbative lattice studies of stealth dark matter scattering are also crucial for dark matter phenomenology, 
but are challenged by a large number of fermion Wick contractions, the signal-to-noise problems, and excited state contamination. Recently, in Ref.~\cite{Brower:2023rqf}, the authors envisioned the use of the Laplacian Heaviside smearing method and the operator projection onto irreducible representations of the lattice octahedral group to overcome some of these challenges, particularly to reduce the excited state contamination. As a first step, they have applied both methods to the calculations of the baryon spectrum including excited states, and for the ground state baryons find consistent results, compared to their previous work in Ref.~\cite{LSD:2014obp} , with reduced statistical errors. 

\begin{figure}
\begin{center}
\includegraphics[width=0.38\textwidth]{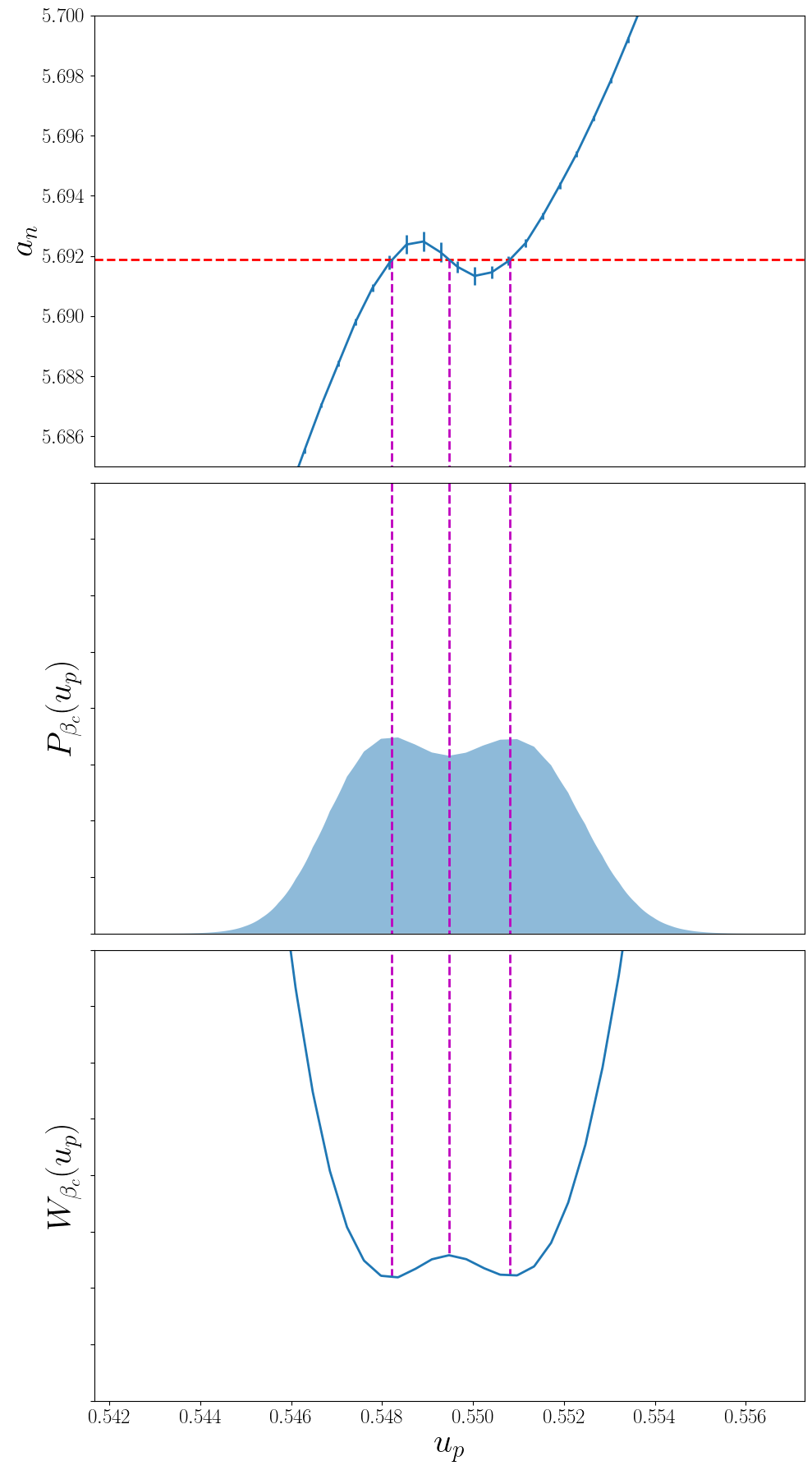}
\hspace{4.0mm}
\includegraphics[width=0.5\textwidth]{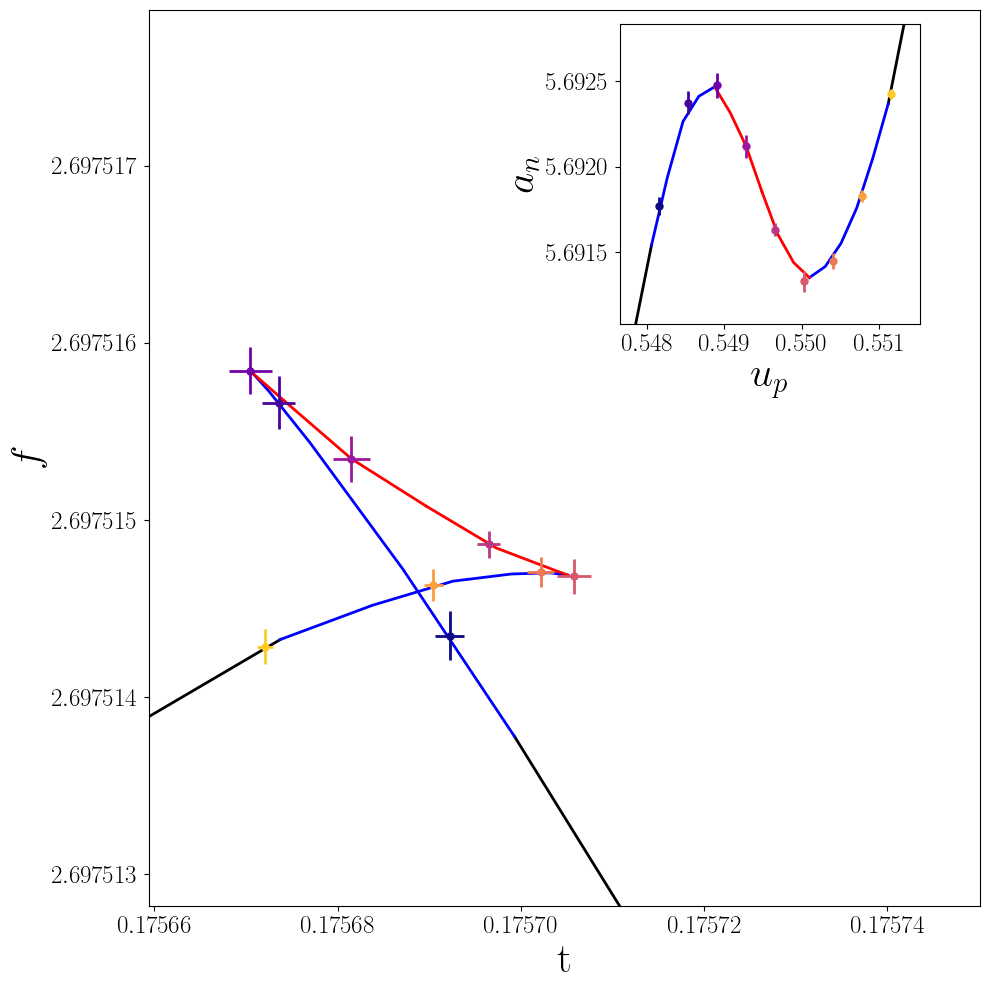}
\caption{%
\label{fig:su3_FT}%
Effective potential and free energy of SU(3) Yang-Mills at finite temperature near the deconfinement transition using the density of states. The plots are taken from Ref.~\cite{Lucini:2023irm}.
}
\end{center}
\end{figure}

Another important aspect of composite dark matter is the first-order finite temperature transition of the underlying dark gauge theories, which may have potential implications for the gravitational waves (GWs) detection in a variety of ongoing and future experiments. 
For instance, NANOGrav collaboration recently reported evidence of a low-frequency stochastic GW background from their $15$ yr data set obtained by pulsar timing array experiments \cite{NANOGrav:2023gor}, where one of the possibilities is that it is a relic of the first-order transitions in the early universe \cite{NANOGrav:2023hvm}. 
First principles lattice field theory calculations will play a crucial role in this regard, but the standard Monte-Carlo importance sampling algorithms are facing technical challenges in sampling the configurations efficiently between two coexisting phases with different energy densities, as well as in measuring the thermodynamic observables with high precision. 
Some recent works have attempted to detour these challenges using the density of states instead of the Monte-Carlo sampling. In particular, focusing on Yang-Mills theories, the Linear Logarithmic Relaxation (LLR) algorithm \cite{Langfeld:2012ah} has been employed to investigate the thermodynamic properties of the (first-order) deconfinement transition in $SU(3)$ \cite{Lucini:2023irm} and $Sp(4)$ \cite{Mason:2023ixv} gauge theories, and the bulk transitions in $SU(N)$ gauge theories with $4\leq N \leq 8$ \cite{Springer:2023hcc} by calculating the density of states accurately. The results are quite promising as seen in \Fig{su3_FT} for the pure $SU(3)$ gauge theory~\cite{Lucini:2023irm}: the left panel shows the inverse (microcanonical) temperature $t=1/a_n$, the plaquette probability distribution $P_{\beta}(u_p)$, and the corresponding effective potential $W_\beta(u_p)$ against the plaquette value $u_{p}$, while the right panel shows the (subtracted) free energy of the micro-states near the deconfinement transition. 
The two-peak structure of $P_\beta$ and the multi-valued nature of the free energy, that are otherwise difficult to see, clearly indicate that the system undergoes a first-order phase transition. 

\section{Other topics}
\label{Sec:others}

{\bf Supersymmetric gauge theories.} Lattice studies of supersymmetric QFTs have potential impacts on BSM physics by providing non-perturbative inputs to supersymmetric extensions of the Standard Model, as well as by providing insights on the nature of quantum gravity via gauge-gravity duality. While simulating supersymmetric QFTs on the lattice is very challenging due to the fact that the lattice regularization breaks the supersymmetry, there has been significant progress in the developments of lattice supersymmetry, focusing on supersymmetric Yang-Mills (SYM) in the four- or lower-dimensional space-time for the last two decades. For a comprehensive review of these recent developments and future prospects, readers are referred to Ref.~\cite{Schaich:2022xgy}. 
An interesting lattice result reported recently shows that the four dimensional $\mathcal{N}=4$ SYM exists in a single non-abelian Coulomb phase for all considered lattice gauge couplings, and the behavior of the static potential is consistent with the holographic prediction, which cannot be seen in perturbation theory \cite{Catterall:2023tmr}. 

Another aspect of the SYM is that a certain sector of the theory is connected to its counterpart of non-supersymmetric field theories through the large $N$ equivalence. For instance, it has been argued that the bosonic sector of a gauge theory coupled to a Dirac fermion in the antisymmetric representation is non-perturbatively equivalent to that of $\mathcal{N}=1$ SYM in the large $N$ limit (orientifold equivalence) \cite{Armoni:2003gp}. 
Together with a number of predictions made in Ref.~\cite{Sannino:2003xe},  this has motivated recent lattice studies of $SU(N)$ gauge theory with one flavor of antisymmetric Dirac fermions. 
The first result with $N=3$ was reported in Ref.~\cite{DellaMorte:2023ylq}, where the meson spectrum has been computed and compared with the analytical predictions derived using an effective field theory approach. Extensions to a larger $N$ are currently underway: some preliminary results and technical challenges were discussed at this conference \cite{DellaMorte:2023sdz}. \\

{\bf Gauge theories with elementary scalars.} A simple extension of the SM is to expand the Higgs sector with additional scalar fields, where the most popular model is the two-Higgs doublet model. Non-perturbative studies are necessary to explore the full parameter space in which certain couplings among the scalar fields could be large. Depending on the size of couplings and the underlying assumptions for the global symmetry, this model exhibits several different phases and the phase transition could be first-order. The latter is of particular interest in many BSM models based on first-order electroweak transition. 
In Ref.~\cite{Catumba:2023jep} an exploratory lattice study on the $SU(2)$ gauge theory coupled to two sets of scalar doublets has been performed at small bare couplings, where the authors first demonstrate that it is possible to tune the couplings such that the electroweak sector of the SM can be reproduced in this model. 
They explored the phase space by computing gauge-invariant observables which can be regarded as an order parameter, and measured the spectrum of the theory in various sectors of the phase space, subject to different global symmetries. The running coupling has also been computed using the gradient flow method and qualitatively behaves as expected from the one-loop perturbative result.\\

{\bf Symmetric mass generation.} Symmetric mass generation is a novel mechanism to give fermions a mass by non-perturbative interactions without breaking chiral symmetry. Realising this mechanism is a crucial step towards building mirror models that target chiral gauge theories in the continuum limit. In recent lattice investigations of the phase diagram of a four dimensional $SO(4)$ invariant Higgs-Yukawa model containing four reduced staggered fermions and a real scalar field, it has been observed that there is a region throughout which the massless and massive symmetric phases appear to be separated by a single continuous phase transition \cite{Butt:2018nkn}. 
More recently, it has been argued that the existence of such phases is related to the cancellation of certain exact 't Hooft anomalies of staggered fermions (promoted to K\"ahler-Dirac fermions), yielding that four reduced staggered fermions, that are equivalent eight Dirac or sixteen Majorana fermions, are required to have a consistent interacting theory allowing four fermion terms, but not fermion bilinear terms \cite{Butt:2021brl,Catterall:2022jky}. 
It is also interesting to realize symmetric mass generation solely from gauge interactions. In fact, in Ref.~\cite{Butt:2021koj}, the authors proposed a lattice model which consists of reduced staggered fermions transforming in the bifundamental representation of a $SU(2)\times SU(2)$ gauge symmetry, and, from numerical investigations of this model, indeed they found a confining phase in which it develops a four fermion condensate but no fermion bilinear. 

\acknowledgments

I would like to thank S.~Catterall, G.~Catumba, A.~Hasenfratz, C.-J.~D.~Lin, B.~Lucini, A.~Lupo, D.~Mason, M.~D.~Morte, D.~Nogradi, C.~Peterson, D.~Schaich, Y.~Shamir and R.~Shrock for providing material in advance of my talk and for fruitful discussions. I also thank F.~Sannino for helpful comments and suggestions on the manuscript.
J.~W.~L. acknowledges support by IBS under the project code, IBS-R018-D1.

\end{document}